\documentclass{aa}

\usepackage{graphicx}
\usepackage{natbib}
\bibpunct{(}{)}{;}{a}{}{,}

\usepackage{amssymb}

\def\ffam{\hbox{$\,.\!\!^{\prime}$}}
\def\ffas{\hbox{$\,.\!\!^{\prime\prime}$}}
\def\ffs{\hbox{$\,.\!\!^{\rm s}$}}
\def\ffm{\hbox{$\,.\!\!^{\rm m}$}}

\begin{document}

\title{Properties and environment of the molecular complex near Holmberg\,IX\thanks{Based on observations carried out with the IRAM Plateau de Bure Interferometer and 30\,m telescope, the 10\,m Heinrich-Hertz-Telescope (HHT), and the NRAO 12\,m telecsope. IRAM (Institut de Radio-Astronomie Millim\'etrique) is supported by INSU/CNRS (France), MPG (Germany) and IGN (Spain). The HHT was operated 
by the Submillimeter Telescope Observatory on behalf of Steward 
Observatory and the Max-Planck-Institut f{\"u}r Radioastronomie. The NRAO (National Radio Astronomy Observatory) is a facility of the National Science Foundation, operated under cooperative agreement by Associated Universities, Inc.} \thanks{Guest User, Canadian Astronomy Data Centre, which is operated by the 
Dominion Astrophysical Observatory for the National Research Council 
of Canada's Herzberg Institute of Astrophysics.} }

\author{F.\,Boone \inst{1} \and N.\,Brouillet \inst{2}  \and S.\,H\"uttemeister \inst{1} \and C.\,Henkel \inst{3} \and J.\,Braine  \inst{2} \and D.\,J.\,Bomans \inst{1} \and F.\,Herpin \inst{2} \and Z.\,Banhidi \inst{1}   \and M.\,Albrecht \inst{1}}
\institute{Astronomisches Institut, Ruhr-Universit\"at, Universit\"atstrasse 150, D-44780 Bochum, Germany \and
Observatoire de Bordeaux, Universit\'e Bordeaux 1 - CNRS,  BP 89, F-33270 Floirac, France \and 
Max-Planck-Institut f\"ur Radioastronomie, auf dem H\"ugel 69, D-53121 Bonn, Germany}
\date{Received 30 April 2004 / Accepted}
\abstract{This paper is aimed at providing  new insight into the nature and origin of the molecular complex situated near the line of sight toward Holmberg\,IX in the M\,81 group of galaxies. The first high resolution CO maps of the complex as well as  single dish $^{13}$CO(1--0), $^{12}$CO(3--2) and millimeter continuum observations and the results of a survey of $^{12}$CO in the region are presented. These data together with the available  H\,{\sc i}, optical and X-ray observations are analyzed to study the properties and environment of the complex. We confirm there is no unobscured massive star formation inside the complex  and from the millimeter constraint on the extinction it must have a low star formation rate or be forming only low mass stars. According to the CO line ratios the abundances and physical conditions could be similar to that of cold gas in spirals. We find from its dynamics (no rotation) and its mass (2--6 million solar masses) that it resembles a massive GMC. Also, re-inspecting  N-body simulations of the M\,81 group and the H\,{\sc i} data we find that it might be located inside the extreme outer disk of M\,81 and be cospatial with the H\,{\sc i} feature known as Concentration\,{\sc i}. The negative result of the CO survey suggests that the complex is unique in this region and calls for a peculiar local formation process. We find that the distribution of the CO emission in the data cube is asymmetrical in a way similar to a cometary object. The optical observations of the nearby supershell MH9/10 suggest the existence of an outflow toward the complex. We consider the possibility that the molecular complex is related to this hypothetical outflow.
 \keywords{ISM: clouds -- Galaxies: individual: M\,81, Ho\,IX -- Galaxies: interactions -- Galaxies: ISM -- Radio lines: ISM -- Submillimeter }}

\maketitle

\defcitealias{1992A&A...262L...5B}{BHB92}

\section{Introduction}
In many cases gravitational interactions between galaxies go 
hand in hand with the loss of matter. Tidal forces act both on
stars and the interstellar gas and dust, leading to tidal tails that may reach linear scales up to 100\,kpc \citep{1994AJ....107...67H}. Some of the ejected interstellar debris may 
cool and condense to molecular clouds, triggering star formation 
that may lead to the formation of tidal dwarf galaxies 
\citep[TDGs; see e.g.][]{2000Natur.403..867B,2001A&A...378...51B,2002A&A...394..823L,2002AJ....123..225W}.

The M\,81 group of galaxies is the nearest one with a significant 
number of gravitationally interacting sources. With its small
distance  \citep[$D$$\sim$3.6\,Mpc according to][]{1994ApJ...427..628F}, high
declination ($\delta$$\sim$69$^{\circ}$) and high galactic latitude
($b^{\rm II}$$\sim$41$^{\circ}$) the M\,81 group is one of the prime 
targets for studies of tidal tails and dense cool intergalactic matter. 

Searching for evidence of star formation in regions of high H\,{\sc i} column
density, \citet[][ hereafter BHB92]{1992A&A...262L...5B} detected CO emission east
of the dwarf irregular Holmberg\,IX. This emission was interpreted in terms
of first direct evidence for the presence of intergalactic molecular clouds
formed out of tidal debris. Thus the molecular complex was supposed to 
provide the missing link between the extended H\,{\sc i} clouds and newly formed 
dwarf irregulars in this region. The mass of the molecular complex was 
estimated to be 10$^{6-7}$\,M$_{\odot}$. A CO $J$=2--1/$J$=1--0 line ratio 
of $\sim$0.5 is consistent with values found in cold ($T_{\rm kin}$$\sim$10\,K) galactic molecular clouds. IRAS (Infrared Astronomical Satellite) data 
allowed \citetalias{1992A&A...262L...5B} to obtain an upper limit to the star formation and ruled out a starburst in this region. Subsequent optical 
observations \citep{1993A&A...273L..15H} did not detect the object, indicating an absence of unobscured individual stars with masses in excess of 
10\,M$_{\odot}$. 

These observations raise important questions about the nature and the origin of the complex. Is it too young to have started any star formation? How did it form if it is intergalactic? Is it related to  Holmberg\,IX? Is it in a tidal arm? In this paper we show that the complex might not be intergalactic but be instead located in the extreme outer disk of M\,81 and that its formation might be related to the nearby unusual supershell MH9/10. We present the first maps of the complex obtained with the Plateau de Bure Interferometer (PdBI) in  $^{12}$CO(1--0) and $^{12}$CO(2--1). We also provide new single dish observations, including $^{13}$CO(1--0), $^{12}$CO(3--2), the millimeter continuum  and  a CO survey in the region surrounding the complex. These observations will be used together with  archival data to study the properties  and  environment of the complex.

\section{Observations}

\subsection{IRAM interferometer CO observations}

The $J$=1--0 and 2--1 lines of $^{12}$CO were measured with the PdBI between April and July 1997 with the configurations 5C, 5D (five antennas) and 4D (four antennas). The 15 m antennas were equipped with dual-band SIS receivers. The spectral correlators were centered at 115.271~GHz and 230.538~GHz respectively, with three correlator units at each frequency: one unit was  centered on the line and two units with bandwidths of 160~MHz were placed on each side of the line to determine the continuum level and calibrate the data (see below). The two centered units have bandwidths of  40~MHz (104~km\,s$^{-1}$) and  80~MHz (104~km\,s$^{-1}$) and resolutions of  0.156~MHz (0.41~km\,s$^{-1}$) and  0.625~MHz (0.82~km\,s$^{-1}$) at 115 and 230\,GHz respectively. The correlator was regularly calibrated by a noise source inserted into the system.

Visibilities were obtained with 20 minute  on-source integrations interspersed by short ($\simeq$5~min) observations of phase and amplitude calibrators toward the quasars 0836+710, 3C454.3, 3C273, 0954+658 and the stellar system MWC349. The data were phase calibrated in the antenna-based mode. From all the observing runs we used only those data for which the residual atmospheric phase jitter was less than or close to 30$\deg$, consistent with a seeing disk of size 0\ffas 6--0\ffas 8  and  a $\simeq$5\% loss of efficiency. The fluxes of the primary calibrators were determined from IRAM 30\,m measurements and were taken as input to derive the absolute flux density scales for our visibilities, estimated to be accurate to $10\%$. The bandpass calibration was carried out using 3C273 and is accurate to better than $5\%$. 

The data reduction was performed using the GILDAS software \citep[e.g.][]{2000irsm.conf..299G}. The short spacings were computed from the 30~m antenna observations of \citetalias{1992A&A...262L...5B} and introduced into the {\it uv}-table using the task SHORT\_SPACE. In this procedure the short spacings are computed from a map built by interpolation of the 30\,m observations, deconvolved from the 30\,m beam and multiplied by the PdBI primary beam. A first guess weight for the short spacings is estimated from the mean weight of the interferometer data in a ring of 1.25 $D/\lambda$ to 2.5 $D/\lambda$ ($D=15$~m) in the {\it uv}-plane. This first guess weight as well as the amplitude of the short spacings can be further adjusted by means of two free parameters. We found that for the flux of the map to be consistent with the flux of the 30\,m data alone scaling factors of 0.6 and 0.8 should be applied to the amplitudes and the weights respectively.

Individual maps with 0\ffas 5 sampling  and  sizes of $256\times 256$ pixels at 115~GHz and $128\times 128$ pixels at 230~GHz were created for each frequency channel.  Natural weighting  and tapering was used at 115~GHz (Gaussian taper of 120~m Full Width at Half Maximum) and  230~GHz (Gaussian of 65~m FWHM). The dirty beam obtained at 115~GHz after tapering is close in size to that at 230~GHz. In the data cube thus obtained, the channel spacing was chosen to be 2.6 km\,s$^{-1}$. The maps were cleaned with the Clark (1980) method and restored with a beam of FWHM $3\ffas 74 \times 3\ffas 31$ and a position angle of $-45.33\deg$ for both lines. Because of the poor signal-to-noise ratio in CO(2--1) the high resolution map (i.e. without tapering) for this line was found difficult to clean and only the map with the same beam as in CO(1--0) is discussed here. The rms noise levels in the cleaned maps are 9 mJy/beam and 20 mJy/beam at 115~GHz and 230~GHz respectively. No continuum emission was detected toward the molecular complex in a 320~MHz wide band down to rms noise levels of 0.5 mJy/beam  and 1 mJy/beam at 115~GHz and 230~GHz respectively. The maps shown are not corrected for primary beam attenuation. The primary beam Full Widths at Half Maximum (FWHM) are equal to   $44^{\prime\prime}$ and  $22^{\prime\prime}$ at 115~GHz and 230~GHz, respectively.

\subsection{Single dish $^{13}CO$}

$^{13}$CO\,($J$=1$-$0) (110\,GHz) line emission was observed  at the position $\alpha_{J2000}=09^h 58^m 0\ffs 7$, $\delta_{J2000}= 69^{\circ}03'23\ffas 0$ with the IRAM 30\,m radiotelescope in May 1998.  We used two SIS receivers in
parallel; the system temperature was $\sim$400\,K on a main beam brightness temperature (T$_{\rm mb}$) scale. Two 1\,MHz
filterbanks and an autocorrelator gave velocity resolutions of 2.6 and
1.3\,km\,s$^{-1}$ respectively.  The half power beamwidth was
$23^{\prime\prime}$.

\subsection{Single dish $^{12}$CO(3--2)}

$^{12}$CO\,($J$=3--2) (345\,GHz)  line emission was observed
 at the position $\alpha_{J2000}=09^h 58^m 1\ffs 7$, $\delta_{J2000}= 69^{\circ}03'18.9\ffas 0$ with the HHT (Heinrich-Hertz-Telescope) in Jan. 1999. The beamwidth 
was $\sim$22$''$, the system temperature was $\sim$900\,K on a T$_{\rm mb}$ scale. The observations were carried out with a dual channel
SIS receiver, two 1024-channel 1 GHz wide Acousto-Optical Spectrometers
(AOS) with a mean channel spacing of $\sim$920\,kHz 
($\sim$0.8\,km\,s$^{-1}$) and a 256-channel filterbank with a channel 
spacing of 250\,kHz ($\sim$0.2\,km\,s$^{-1}$).  A main beam brightness temperature was 
established using a beam efficiency of 0.5 and a forward hemisphere 
efficiency of 0.9 \citep[see e.g.][]{2002A&A...389..589M}.

\subsection{Single dish millimeter continuum}
We observed the complex at 1.2~mm with the MAMBO bolometer array at the IRAM 30\,m antenna in September/October 2003. We used an on-off mode to achieve the highest sensitivity at the position  $\alpha_{J2000}=09^h 58^m 1\ffs 7$, $\delta_{J2000}= 69^{\circ}03'23\ffas 0$. The data were reduced using the NIC package of the GILDAS software. The noise level is 0.3\,mJy, no signal was detected.

\subsection{Single dish $^{12}$CO survey}

Several positions around the complex (for details, see Sect.\,\ref{sec:survey}) were observed with the IRAM 30\,m and the NRAO 12\,m telescope. The 30\,m measurements were made in the CO(1--0) line in Feb. 1998 in a similar way as described in Sect.\,2.2. 

In Nov. 1992 and Feb. 98 the   $^{12}$CO(1--0) (115\,GHz) line emission was observed with the  NRAO 12\,m telescope at Kitt Peak. Two SIS receivers were used in parallel, the system temperature was in the range  350-600\,K (T$_{\rm R}^*$). A 256-channel filterbank with a channel spacing of 1\,MHz and the autocorrelator of 600\,MHz bandwidth and  a resolution of 781\,kHz were used as backends. The beam FWHM is 55$''$.

In Feb. 1998 the  $^{12}$CO(2--1) (230\,GHz) line emission was observed with the 8-beam array at the NRAO 12\,m. The same 256-channel filterbank was used as well as an hybrid spectrometer of 300\,MHz bandwidth. The system temperature was in the range 800-1400\,K. The FWHM beam size is 27$''$.

\subsection{Optical data}

The optical data were retrieved from the Canadian Astronomy Data Centre (CADC) archive. The B, R, and H$\alpha$ images of the
field near Holmberg\,IX were taken by  Pakull and Mirioni with the OSIS instrument at the 3.6\,m Canada-France-Hawaii Telescope (CFHT) and were published in conference proceedings \citep{pakull2002,2003RMxAC..15..197P}. 
Data reduction was perfomed using IRAF\footnote{
IRAF is distributed by the National Optical Astronomy Observatories,
    which are operated by the Association of Universities for Research
    in Astronomy, Inc., under cooperative agreement with the National
    Science Foundation.} software \citep{Tody93}. 
The data were bias subtracted and flatfielded using the 
standard methods.  Special care was taken to align the images 
to a common reference using stars present on the B, R, and H$\alpha$ 
images. The continuum subtracted H$\alpha$ image was produced 
subtracting an intensity scaled R image, as described in
\citet{Bomans97}.
Flux calibration of the continuum subtracted H$\alpha$ image was 
performed by using the fluxes of three isolated H\,{\sc ii} regions present 
in both our image and the data of \citet{Miller94,1995ApJ...446L..75M}.

\section{Properties of the complex}

\begin{figure*}
\begin{center}
\resizebox{17cm}{!}{\rotatebox{0}{\includegraphics{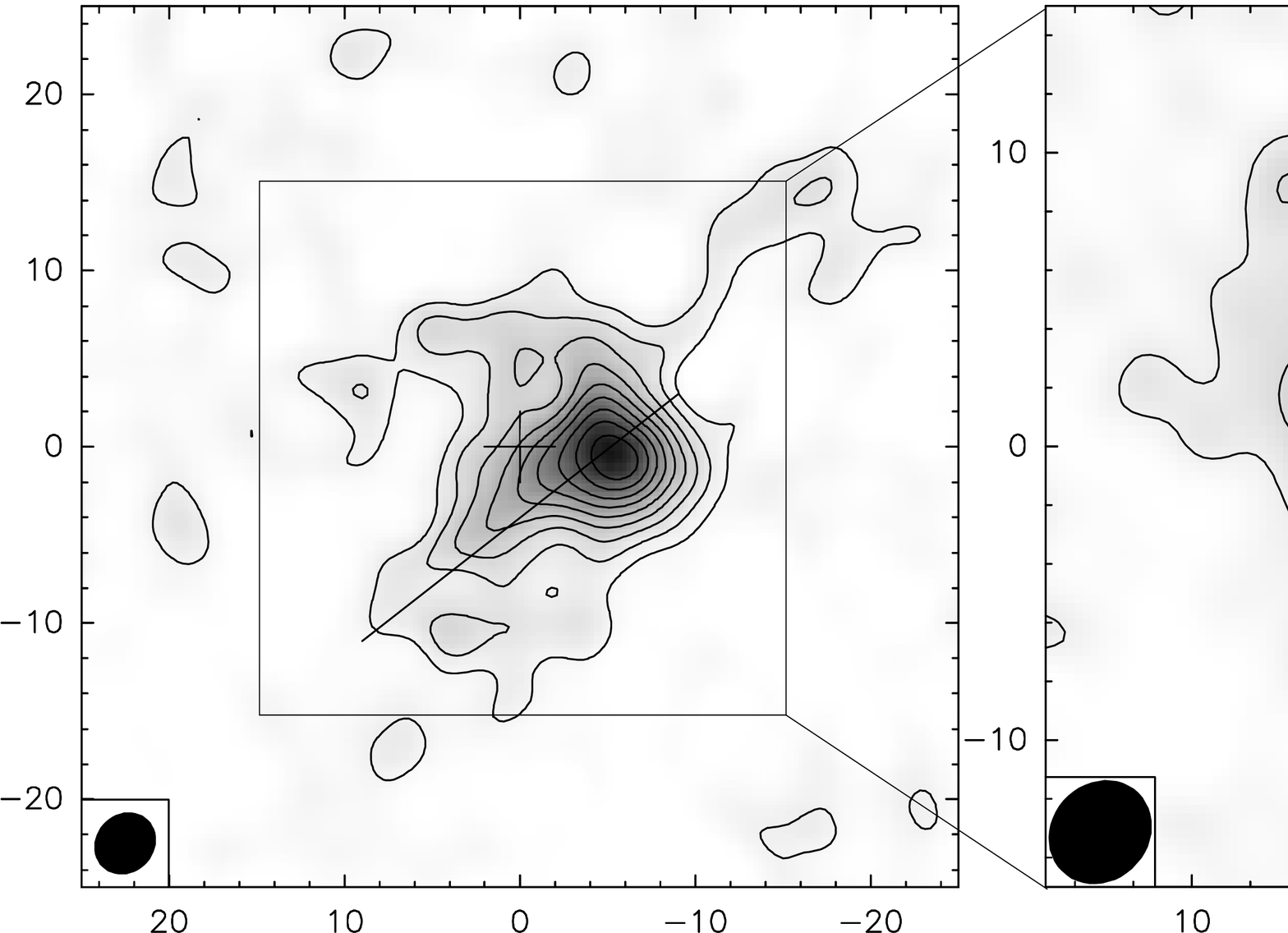}} \rotatebox{0}{\includegraphics{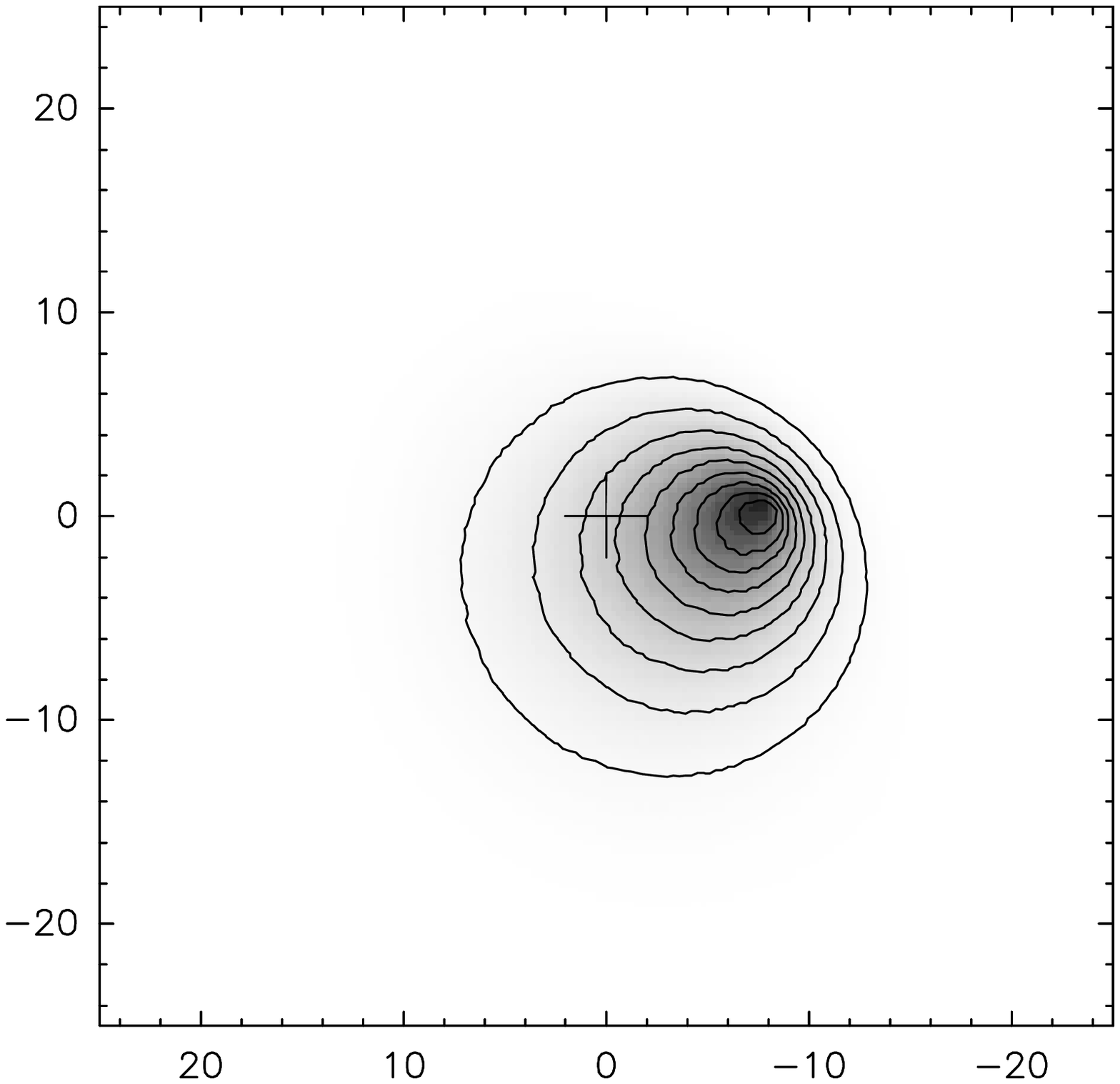}} }
\caption{$^{12}$CO(1--0) (left), $^{12}$CO(2--1) (middle) and model (right) integrated maps from -45~km\,s$^{-1}$ to -10~km\,s$^{-1}$ ($V_{\rm LSR}$). The axes give the  R.A. and Dec. offsets in arcseconds with respect to the phase center ($\alpha_{J2000}=09^h 58^m 1\ffs 7$ and $\delta_{J2000}= 69^{\circ}03'23\ffas 0$) indicated by a cross. The $^{12}$CO(1--0) contours (left panel)  go from 340~mJy/beam~km\,s$^{-1}$ ($4\,\sigma$) to 1.7~Jy/beam~km\,s$^{-1}$ ($20\,\sigma$) by steps of 170~mJy/beam~km\,s$^{-1}$ ($2\,\sigma$). The $^{12}$CO(2--1) contours (central panel) go from 0.8~Jy/beam~km\,s$^{-1}$ ($4\,\sigma$) to 3.2~Jy/beam~km\,s$^{-1}$ ($16\,\sigma$) by steps of 0.4~Jy/beam~km\,s$^{-1}$ ($2\,\sigma$). The segment correponding to the position-velocity diagram of Fig.\,\ref{fig:pvplot} is shown on the CO(1--0) map. The cone model (right panel) is discused in Sect.\,\ref{sec:disc}}
\label{fig:maps}
\end{center}
\end{figure*}

\begin{figure*}
\resizebox{17cm}{!}{\rotatebox{-90}{\includegraphics{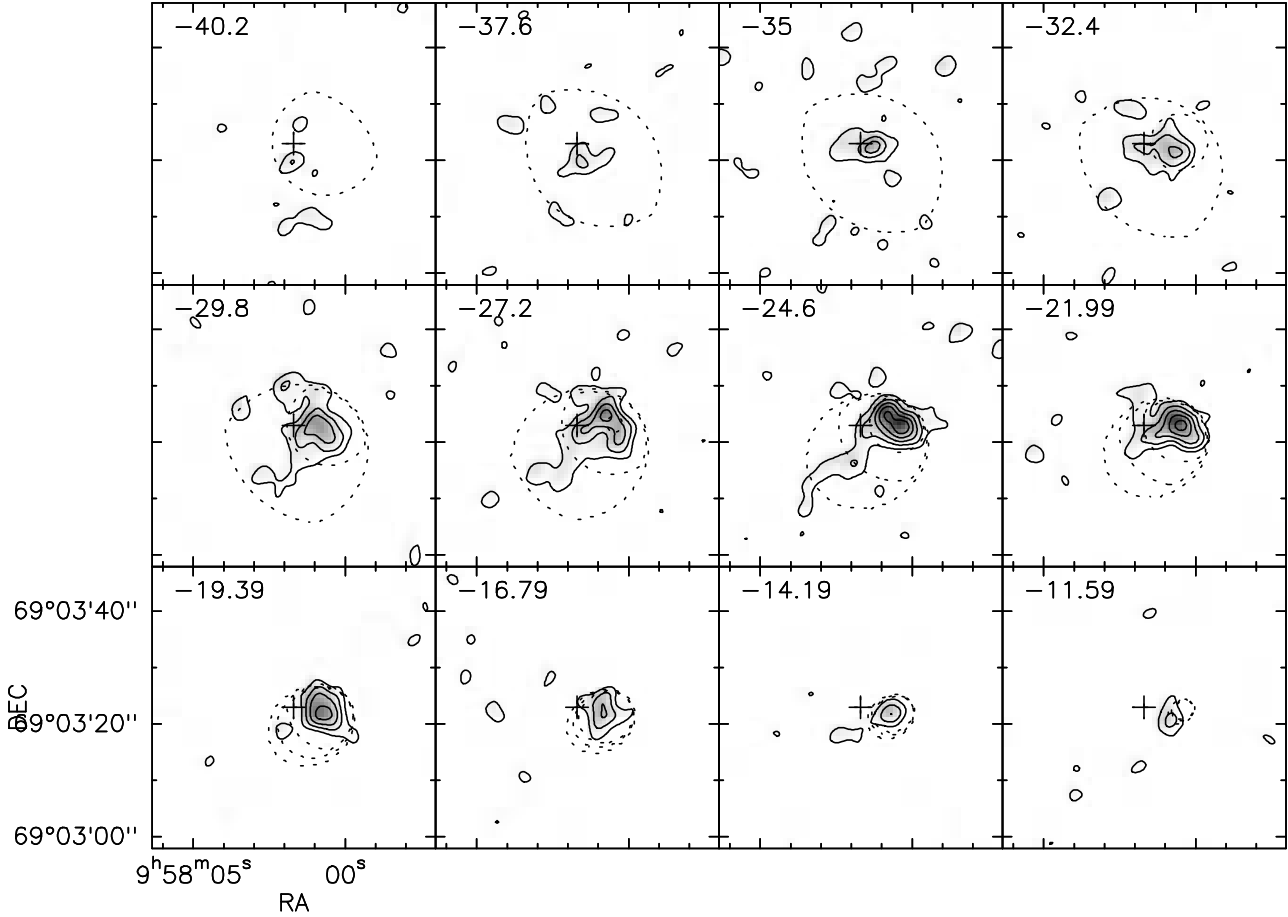}}}
\caption{$^{12}$CO(1--0) velocity-channel maps observed with the IRAM interferometer and the IRAM 30\,m antenna for the short spacings. The resolution (FWHM of the beam) is  $3.74"\times 3.31"$ (PA=$-45.5\deg$). Velocity channels range from $V_{\rm LSR}=-40.2$ to $-11.6$~km\,s$^{-1}$ by steps of 2.6\,km\,s$^{-1}$. The phase center is marked by a cross in each map. The contours begin at 27 mJy/beam (i.e. $3\,\sigma$) and are spaced by 18 mJy/beam ($2\,\sigma$), the highest contour is at 99 mJy/beam ($11\,\sigma$). The dotted lines represent contours of a cone model, the levels are at 1/30, 4/30 and 7/30 of the maximum of the model data cube (The maximum of individual maps is decreasing when going from $-11$ to $-40$\,km\,s$^{-1}$ and the higher levels disappear accordingly).}
\label{fig:chanmaps}
\end{figure*}

\begin{figure}
\resizebox{\hsize}{!}{\rotatebox{0}{\includegraphics{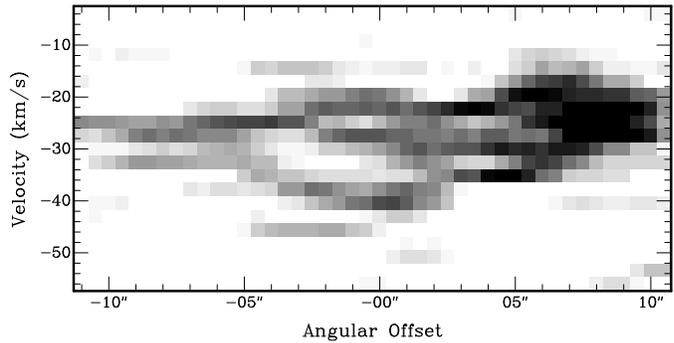}}}
\caption{$^{12}$CO(1--0) position-velocity diagram along the segment drawn on Fig.\,\ref{fig:maps} (left panel). The right end of the angular offset axis corresponds to the N-W end of the segment.}
\label{fig:pvplot}
\end{figure}

\subsection{Morphology and dynamics}\label{sec:morph}

The overall $^{12}$CO(1--0) and  $^{12}$CO(2--1) integrated maps are shown Fig.\,\ref{fig:maps}. The  $^{12}$CO(1--0) maximum is 1.94~Jy/beam~km\,s$^{-1}$ at $\alpha_{J2000}=09^h58^m0\ffs 71$, $\delta_{J2000}= 69^{\circ}03'22\ffas 6$. The integrated emission  shows an abrupt edge on its N-W side and a smoother intensity gradient toward the S-E, with however a low-level filament to the N-W. 

The  $^{12}$CO(2--1) maximum is  3.57~Jy/beam~km\,s$^{-1}$ and is located at the same position as the CO(1--0) peak. The spatial distribution of the integrated emission looks also elongated in the N-W to S-E direction. Some extended emission at the 5\,$\sigma$ level appears on the N-E side of the main elongated shape. This could be a hint for the presence of warm diffuse gas at the periphery of the complex.

Figure \ref{fig:chanmaps} displays the 12 channel maps of the $^{12}$CO(1--0) line data cube showing emission over $3\,\sigma$. They cover a velocity range of $\sim$30 km\,s$^{-1}$ from $V_{\rm LSR}=-40.2$~km\,s$^{-1}$ to $-11.6$~km\,s$^{-1}$  with a resolution of 2.6 km\,s$^{-1}$. Most of the emission stronger than $3\,\sigma$  is found in a circular region of 20'' ($\sim$350~pc) in diameter. 

Noticeable features are: (1) A compact region with high-level emission about $\sim$6'' west from the phase center. It is most clearly seen in channels $-24.6$ to $-19.4$ km\,s$^{-1}$. (2) Low-level extended emission in the form of filaments (channels from $-29.8$ to $-22$~km\,s$^{-1}$) and clumps (mostly in channels from $-37.6$ to  $-32.4$~km\,s$^{-1}$).  Although the clumps are close to the 3\,$\sigma$ level, the fact that they are mostly seen in three adjacent channels  indicates that they are hinting at real low level emission. Most of the clumps have  velocity widths of one channel only (2.6 km\,s$^{-1}$). 

As for the global properties of the distribution, we note: (1) There is no strong evidence of rotation at a velocity higher than 15~km\,s$^{-1}$ over distances greater than 10'' (i.e. 175~pc), nor of disk-like structure. (2)  Although Fig.\,\ref{fig:pvplot} might suggest the presence of a cavity that appears as a loop in the diagram  (between 0 and $-5''$ angular offset), it is not clear that this is caused by stellar activity like bubble or outflow structures \citep[BHB92,][]{1993A&A...273L..15H}.

 The global cloud morphology noted in the integrated maps (Fig.\,\ref{fig:maps}) and the filaments seen in the channel maps (Fig.\,\ref{fig:chanmaps}) are reminiscent of a  cometary object pointing toward the N-W. Even though giant molecular clouds (GMC) in general are expected to be assymmetric, this shape could give hints on the origin and the nature of the complex. This point will be discussed in Sect.\,\ref{sec:disc}.

\subsection{Mass of the complex}\label{sec:mass}

Summing up the CO(1--0) data in a polygon enclosing the emission in the velocity interval $[-45,-10]$~km\,s$^{-1}$ gives a total flux  of $18$~Jy\,km\,s$^{-1}$. The greatest uncertainty comes from the flux calibration that is estimated to be  $\sim$$\pm$20\%.
By assuming a linear correlation between the H$_2$ column density and the CO(1--0) line intensity, this flux can be converted into an H$_2$ mass. For a distance to the object of $D=3.6$~Mpc we get:
\begin{equation}
M_{{\rm H}_2}=[2.7\pm 0.6] \left[\frac{X}{3\times 10^{20} }\right]\times 10^6~{\rm M}_{\odot},
\end{equation}
where $X$ is the conversion factor  in cm$^{-2}$\,(K\,km\,s$^{-1}$)$^{-1}$ relating the H$_2$ column density to the CO(1--0) line intensity.  

The value of $X$ for the Galactic disk is estimated in the range 2-$3\times 10^{20}$cm$^{-2}$(K~km\,s$^{-1}$)$^{-1}$ \citep{1986A&A...154...25B, 1991ARA&A..29..581Y, 1991IAUS..146..235S}. Estimates of $X$ in extragalactic sources have been addressed in numerous references and the conversion factor is thought to depend mostly on the radiation field, the gas-to-dust ratio and the metallicity \citep[e.g.][]{1989ApJ...344L..61D, 1996PASJ...48..275A, 1999ApJ...513..275B, 2001mhs..conf..293I}. There does not seem to be any strong radiation source within the complex but the gas-to-dust ratio of the complex  discussed in Sect.\ref{sec:dust} could be higher than the galactic value (150). In this case the conversion factor would also be higher. 

The metallicity is expected to be lower than in central regions of galaxies. Indeed, irrespective of whether the complex is situated in the disk of M\,81 or outside of it, it is at a large distance from the nucleus and the metallicity is known to decrease with galactocentric distance. $X$ is therefore probably higher than the galactic disk value and $2.7 \times 10^6~{\rm M}_{\odot}$ can be seen as a lower limit for the mass of the complex. More precisely if we extrapolate the radial distribution of the [OII]+[OIII] to H$\beta$ ratio measured by \citet{1992ApJ...390L..73Z}  to the deprojected radius of the complex, i.e. 23\ffam 1, we get a value of 0.7 which corresponds, from the calibration of \citet{1984MNRAS.211..507E}, to an abundance of 12+log(O/H)$\simeq$8.6. If we take the relationship between $X$ and the metallicity proposed by \citet{2001mhs..conf..293I} namely log$(X)=-2.7\,$log(O/H)+11.6 we get $X\simeq 6.0\times 10^{20}$ which would mean a mass twice as large: $5.4 \times 10^6~{\rm M}_{\odot}$. We note that this estimate is subject to large uncertainties: the conversion of the  [OII]+[OIII] to H$\beta$ ratio into oxygen abundance is uncertain and the expression of $X$ as a function of metallicity is empirical and is based on the data of a few objects only. To summarize, the mass of the complex is estimated to be in the range  $2-6 \times 10^6~{\rm M}_{\odot}$. This mass is comparable to that of the largest GMCs also called Giant Molecular Associations (GMAs) in the  Galaxy.

\subsection{Physical conditions}
\begin{figure}
\resizebox{\hsize}{!}{\rotatebox{0}{\includegraphics[trim=30 50 1 1, clip]{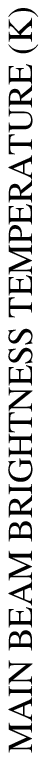}}}
\caption{The $^{13}$CO(1--0) and $^{12}$CO(3--2) lines from the molecular complex. The $^{12}$CO(1--0) line divided by 6 and 2.5, respectively, is overlaid as a dashed line.}
\label{fig:30m98}
\end{figure}

\begin{table}
\caption{Integrated $^{12}$CO(1--0)/$^{13}$CO(1--0) and $^{12}$CO(3--2)/(1--0) line ratios for different spectral windows.}
\begin{tabular}{rccc}
  & window & $^{12}$CO/$^{13}$CO & (3--2)/(1--0) \\
 & (km\,s$^{-1}$) & & \\
center of line & $[-32$, $-18]$ & 7.1$\pm$1.8  & 0.26$\pm$0.06 \\
wider center   & $[-37$, $-13]$ & 10.4$\pm$3.2 & 0.16$\pm$0.05 \\
whole line     & $[-42$, $-8]$  & 14.3$\pm$5.8 & 0.19$\pm$0.06 \\
blue wing      & $[-37$, $-23]$ & 14.7$\pm$7.3 & 0.11$\pm$0.06 \\
red wing       & $[-27$, $-13]$ & 6.3$\pm$1.9  & 0.24$\pm$0.08 \\

\end{tabular}
\label{tab:ratio}
\end{table}
The  integrated $^{12}$CO(2--1)/(1--0) line intensity ratio computed 
from the PdBI data with the zero spacings from the IRAM 30\,m  is 
in the range 0.4-1.6, and considering the uncertainties most of the 
pixels are consistent with a value of 0.6-0.7. The highest values 
are reached at the periphery of the complex, which can be interpreted 
in terms of enhanced kinetic temperatures or low optical depths.
 The integrated  $^{12}$CO(1--0)/$^{13}$CO(1--0) and  
$^{12}$CO(1--0)/(3--2) ratios computed from the single dish data shown in 
Fig.\,\ref{fig:30m98} are given in Table \ref{tab:ratio} for different spectral windows. 

The Large Velocity Gradient (LVG) approximation can be used to 
           compute the radiative transfer within a cloud of uniform
           density and temperature. This is appropriate for a zero order
           simulation of the physical conditions of the complex. 
With a column density $N$(H$_2$) = (0.5--1.0)$\times$$10^{22}$\,cm$^{-2}$ obtained from the mass and size of the complex, a CO abundance [CO/H$_2$]$=8\times 10^{-5}$ \citep{2003ApJ...586..891B} and varying the abundance ratio [$^{12}$CO/$^{13}$CO] in the interval $[40,\,100]$ \citep{1994ARA&A..32..191W} and the line ratios in the intervals $[0.6,\,0.7]$ for $^{12}$CO(2--1)/(1--0),  $[7,\,13]$ for $^{12}$CO/$^{13}$CO and $[0.11,\,0.26]$ for 
$^{12}$CO(3--2)/(1--0) our LVG model gives temperatures in the range 10--30\,K and densities in the range 200--700\,cm$^{-3}$, area filling factors of order $f_{\rm a}$$\sim$0.02 and a velocity gradient of $\sim$10\,km\,s$^{-1}$\,pc$^{-1}$. This velocity gradient is larger than the global velocity gradient obtained from the cloud size ($\sim$300\,pc) divided by the line width  ($\sim$30\,km\,s$^{-1}$), i.e. 0.1\,km\,s$^{-1}$\,pc$^{-1}$. This is probably due to a high dispersion in the central regions of the 
complex.

It is interesting to note that the line ratios (and therefore the 
solution they admit in terms of abundances and physical conditions) are 
similar to those found in cold molecular gas in the inner disks of spiral 
galaxies and that crucial parameters, like size of the complex,
velocity dispersion and volume averaged densities follow the relations 
established by \citet{1981MNRAS.194..809L} ane \citet{1985prpl.conf...81M}.

\subsection{Dust}\label{sec:dust}

The non-detection of the 1.2~mm continuum implies an upper $3\,\sigma$ flux limit of 1~mJy in an 11$''$ beam. With a mass absorption coefficient $\kappa_{\rm 1.2mm}=0.035$~m$^2$kg$^{-1}$ \citep{2002MNRAS.335..753J} and an assumed dust temperature of 10\,K the limit corresponds to a dust mass of $M_{\rm d}=17\times 10^3\,{\rm M}_{\odot}$. The mass being lower for higher temperatures, this is an upper limit. The H$_2$ mass deduced from the CO seen in the same beam at the same position is $M_{{\rm H}_2}=1.0\times 10^6~{\rm M}_{\odot}$ with the standard conversion factor $X=3\times 10^{20}$\,cm$^{-2}$\,(K\,km\,s$^{-1}$)$^{-1}$ which was already introduced in Sect.\,\ref{sec:mass} as a lower limit for outer disks. The corresponding H\,{\sc i} flux is 0.1~mJy in a 12$''$ beam  \citep[data from ][]{1996AJ....111..735A}. Assuming a uniform distribution this corresponds to an H\,{\sc i} mass of $M_{\rm H{I}}=0.5\times 10^6~{\rm M}_{\odot}$ in a 11'' beam. Taking into account the contribution of He as being 40\%  of the H mass (28\% of the ISM mass), this leads to a gas-to-dust ratio lower limit $M_{\rm g}/M_{\rm d}\ge 123$. This lower limit does not exclude the standard galactic value of 150. 

The 3\,$\sigma$ upper limit of 1\,mJy on the dust emission allows to  derive an upper limit for the visual extinction. In the optically thin limit $A_{\rm V}$  is given by:
\begin{equation}
A_{\rm V}=1.086\times S_{\rm 1.2mm}\left(B_{\rm 1.2mm}(T)\,\Omega_{\rm mb}\right)^{-1}\frac{\kappa_{\rm V}}{\kappa_{\rm 1.2mm}},
\end{equation}
where $S_{\rm 1.2mm}$ is the flux, $B_{\rm 1.2mm}(T)$ is the black body brightness and $\Omega_{\rm mb}$ is the antenna main beam.
Estimates of the ratio ${\kappa_{\rm V}}/{\kappa_{\rm 1.2mm}}$ \citep{1998A&A...329L..33K, 2003A&A...399L..43B} imply that it is most probably lower than $1.6\times 10^5$. Then, for a dust temperature of 10\,K and a beamsize of 11$''$, the 1\,mJy upper limit yields: $A_{\rm V}$$<$5\ffm 5.

\subsection{Star formation}\label{sec:sf}

\begin{figure}
\resizebox{\hsize}{!}{\rotatebox{-90}{\includegraphics[clip]{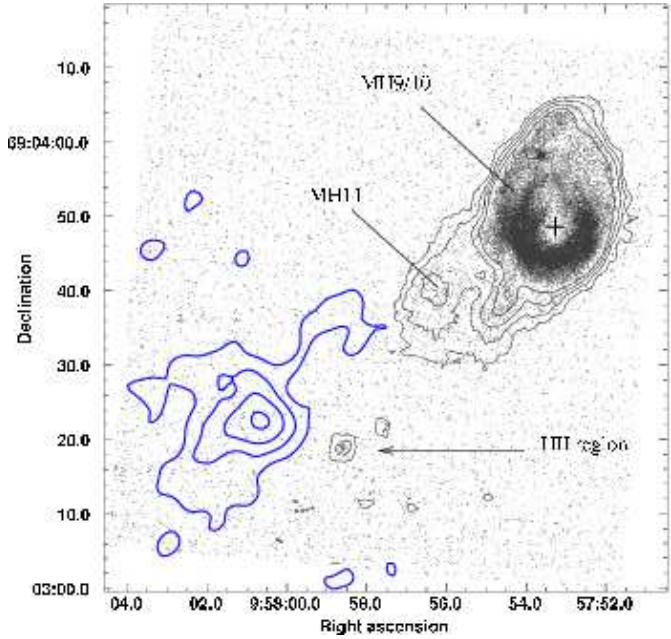}}}
\caption{H$\alpha$ emission (grey scale) from  CFHT observations. The thin contours correspond to the low level H$\alpha$ (lowest contour at 3\,$\sigma$ with a contour smoothness of 10), the thick contours to the CO(1--0) levels (the lowest at 4\,$\sigma$). The cross in the shell shows the Chandra position of X-9. MH9/10 is a large supershell  \citep{1995ApJ...446L..75M}. The MH11 nebula was catalogued as a separate H{\sc ii} region in  \citet{Miller94}.}
\label{fig:Ha}
\end{figure}

\begin{figure}
\resizebox{\hsize}{!}{\rotatebox{-90}{\includegraphics[clip]{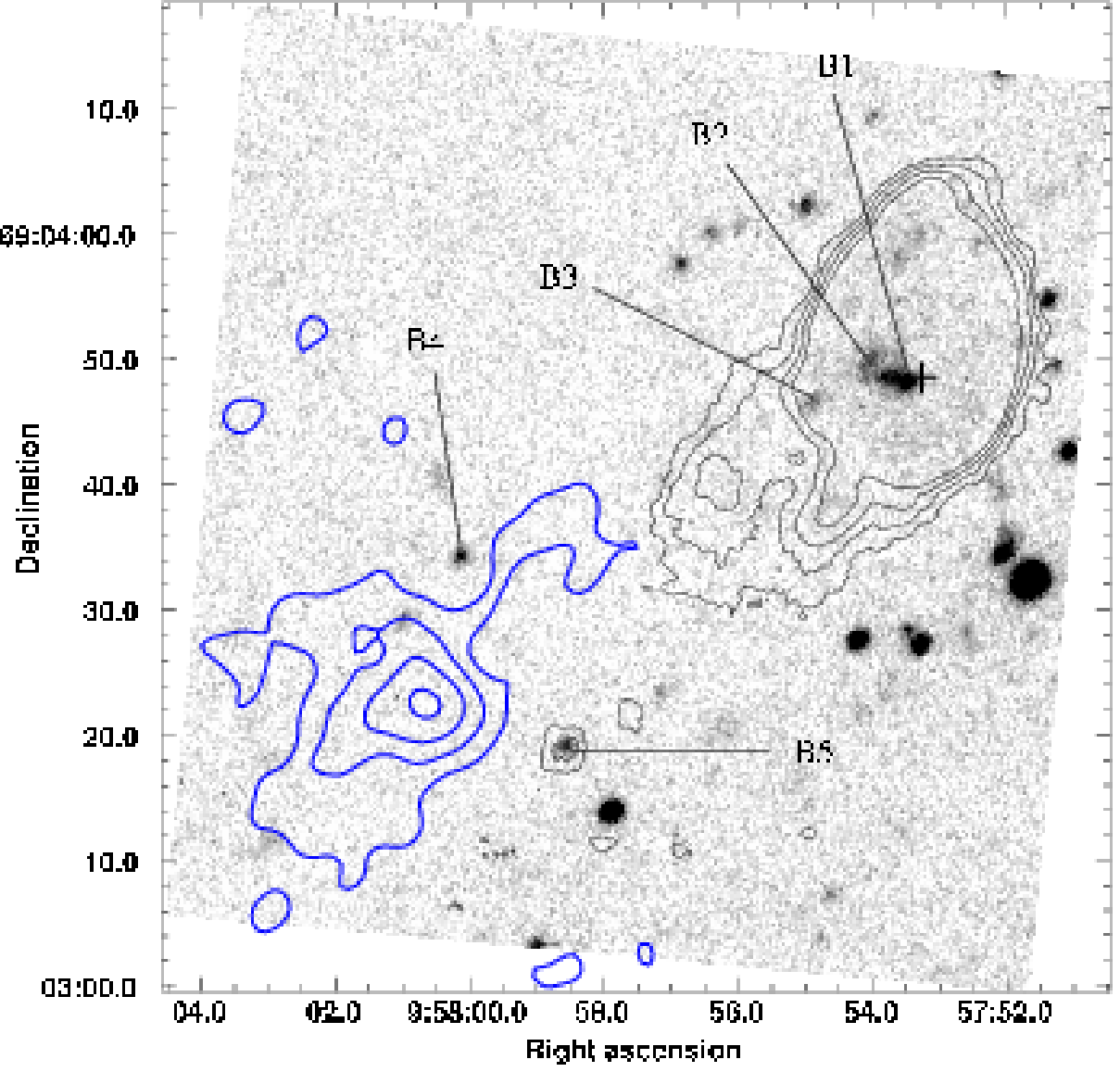}}}
\caption{B band image (grey scale) from  CFHT. The thin contours correspond to the low level H$\alpha$ (lowest contour at 3\,$\sigma$), the thick contours to the CO(1--0) levels (the lowest at 4\,$\sigma$). The cross in the shell shows the Chandra position of X-9.}
\label{fig:B}
\end{figure}

\begin{figure}
\resizebox{\hsize}{!}{\rotatebox{-90}{\includegraphics[clip]{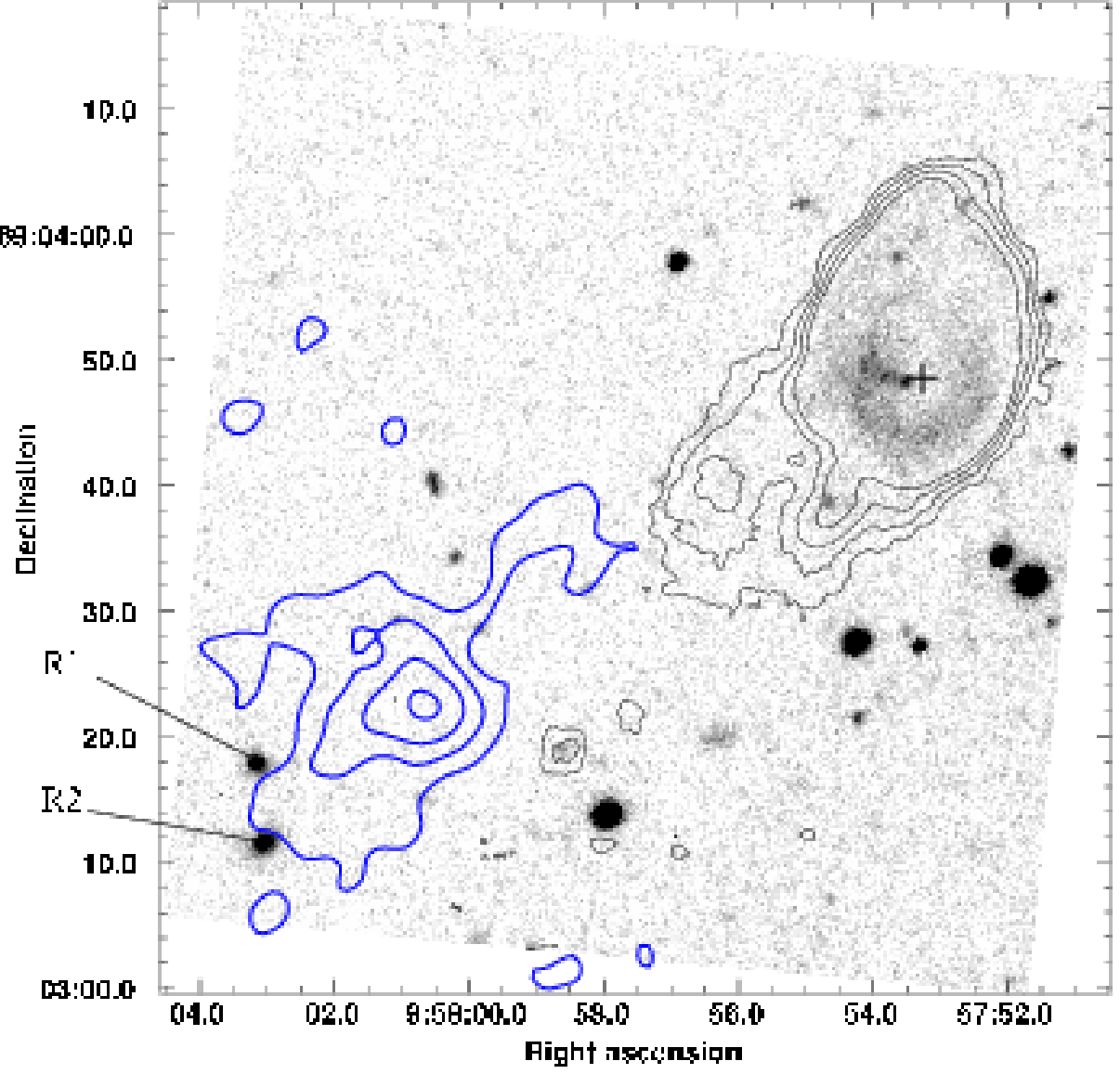}}}
\caption{R band image (grey scale) from  CFHT. The thin contours correspond to the low level H$\alpha$ (lowest contour at 3\,$\sigma$), the thick contours to the CO(1--0) levels (the lowest at 4\,$\sigma$). The cross in the shell shows the Chandra position of X-9.}
\label{fig:R}
\end{figure}

The H$\alpha$ frame from the CFHT does not show any emission  within the 4\,$\sigma$ CO contour as shown in Fig.\,\ref{fig:Ha}. The r.m.s. level of the continuum subtracted H$\alpha$ image is at $10^{-17}$\,erg\,cm$^{-2}$\,s$^{-1}$. To derive an upper limit to the star formation rate we use SFR$=5\times 10^{-8}L_{\rm H\alpha}/L_{\odot}\, {\rm M}_{\odot}$\,yr$^{-1}$ \citep{1986PASP...98....5H} and we correct for the galactic extinction  $A_{\rm R}$$=$0\ffm 210. Without considering any intrinsic extinction this leads to: 
\begin{equation}
\mbox{SFR}<10^{-6} \ \mbox{M}_{\odot} \  \mbox{yr}^{-1}.
\end{equation}
This is an upper limit on the unobscured SFR. When assuming the maximum extinction allowed by the millimeter observation we get:
\begin{equation}
\mbox{SFR}<10^{-4} \ \mbox{M}_{\odot} \  \mbox{yr}^{-1}.
\end{equation}
 Thus an SFR of $10^{-3} {\rm M}_{\odot}$\,yr$^{-1}$ --  expected for  $10^{6}\,{\rm M}_{\odot}$ of molecular gas -- can be excluded for the Salpeter Initial Mass Function (IMF) implicit in the relation of \citet{1986PASP...98....5H}.

While no  H$\alpha$ emission is seen within the complex we identify a new small H\,{\sc ii} region on its western side about 10'' from the CO peak with the coordinates $\alpha_{J2000}=09^h 57^m 58\ffs 6$, $\delta_{J2000}= 69^{\circ}03'19\ffas 0$. At $D$=3.6\,Mpc the H$\alpha$ luminosity of this object corrected for  galactic extinction is $L_{{\rm H}\alpha}=1.7\times 10^{35}$\,erg\,s$^{-1}$. Making the standard assumption that the total number of Lyman continuum photons is equal to the number of H$\alpha$ photons times the ratio of the recombination coefficient for ``Case B'' to that of H$\alpha$  \citep{1989agna.book.....O} we find the number of Lyman continuum photons emitted per sec: $N_{\rm Lyc}\sim 8\times 10^{46}$ s$^{-1}$. From the calibrations of \citet{1973AJ.....78..929P} this object is too faint to contain any O-type star but 3 B0{\sc V} stars could produce the computed ionising flux. This object is also seen in the blue band (Fig.\,\ref{fig:B} object B5) which is compatible with a small association of massive stars, arguing  against a planetary nebula. We also note another blue object on the North of the complex (B4). The two red objects mentioned in \citet{1993A&A...273L..15H} and interpreted as possible background galaxies are also seen in the R band image (Fig.\,\ref{fig:R} objects R1 and R2). The fact that they lie outside the 4\,$\sigma$ level of the CO(1--0) emission supports their interpretation as background galaxies.

\section{Environment}\label{sec:env}
\begin{figure*}
\centering
\resizebox{\hsize}{!}{\rotatebox{0}{\includegraphics{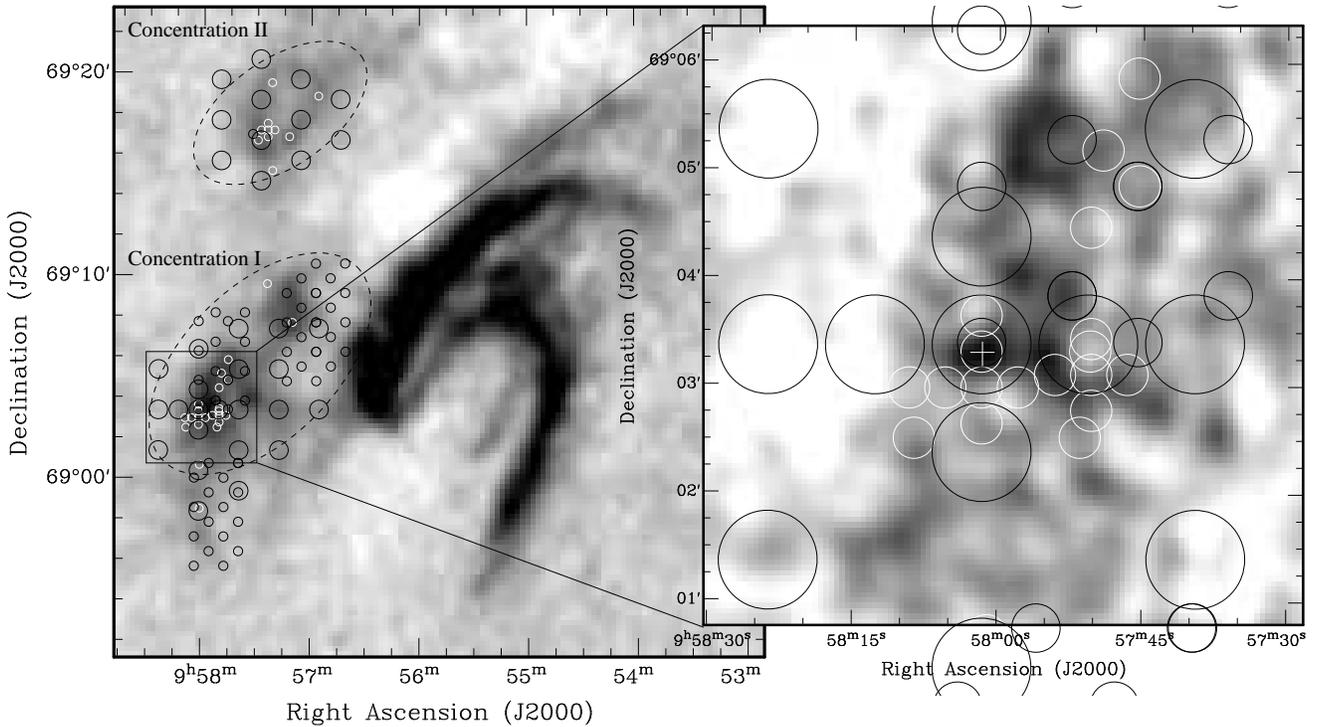}}}
\caption{Observed positions of the CO survey superposed onto a map of H\,{\sc i} emission from \cite{1994Natur.372..530Y} (left) and from \cite{1996AJ....111..735A} (right). The white cross on the right panel shows the position of the complex. The circles represent the beams of the IRAM 30\,m (white) and the Kitt Peak telescope (black; large circles 3\,mm; small circles 1\,mm). The dashed ellipses show the locations of Concentration\,{\sc i} and Concentration\,{\sc ii}.}
\label{fig:1}       
\end{figure*}

\subsection{Search for other CO complexes}\label{sec:survey}

\begin{table}
\caption[]{Positions observed in CO(1--0) with the IRAM 30 meter telescope.
Column 3 gives the rms noise level for each spectrum in milliKelvins (main beam brightness temperature) and
col. 4 the $1\,\sigma$ uncertainty in the CO integrated intensity in K km\,s$^{-1}$.}
\label{tab30m}
\begin{center}
\begin{tabular}{llll}
\hline
RA & Dec & rms & $\Delta$I$_{\rm CO}$ \\
B1950 & B1950 & mK & K km\,s$^{-1}$ \\
\hline
9:53:05.0 & 69:21:30 & 20.3 & 0.15  \\
9:53:05.0 & 69:22:00 & 20.6 & 0.15  \\
9:53:05.0 & 69:22:30 & 19.5 & 0.14  \\
9:53:05.0 & 69:23:00 & 20.1 & 0.14  \\
9:53:05.0 & 69:23:30 & 20.5 & 0.15  \\
9:53:05.0 & 69:24:00 & 20.9 & 0.15  \\
9:53:20.0 & 69:30:00 & 13.5 & 0.10  \\
9:53:22.0 & 69:30:00 & 21.9 & 0.16  \\
9:53:18.0 & 69:30:00 & 20.4 & 0.15  \\
9:53:22.0 & 69:29:30 & 14.5 & 0.10  \\
9:53:20.0 & 69:29:30 & 22.9 & 0.17  \\
9:53:20.0 & 69:30:30 & 19.7 & 0.14  \\
9:53:18.0 & 69:30:30 & 14.3 & 0.10  \\
9:53:18.0 & 69:31:00 & 13.8 & 0.10  \\
9:53:16.0 & 69:31:00 & 13.3 & 0.10  \\
9:53:27.0 & 69:31:18 & 11.3 & 0.08  \\
9:53:29.0 & 69:31:18 & 14.8 & 0.11  \\
9:53:27.0 & 69:31:48 & 16.2 & 0.12  \\
9:53:25.0 & 69:31:18 & 15.9 & 0.11  \\
9:53:27.0 & 69:30:48 & 15.3 & 0.11  \\
9:53:55.9 & 69:18:20 & 22.7 & 0.16  \\
9:53:55.9 & 69:18:40 & 22.7 & 0.16  \\
9:53:55.9 & 69:19:00 & 21.6 & 0.16  \\
9:53:55.9 & 69:19:20 & 21.8 & 0.16  \\
9:53:55.9 & 69:19:40 & 21.6 & 0.16  \\
9:53:55.9 & 69:20:00 & 22.8 & 0.16  \\
9:53:54.8 & 69:19:50 & 14.1 & 0.10  \\
9:53:54.8 & 69:19:30 & 13.6 & 0.10  \\
9:53:54.8 & 69:19:10 & 13.0 & 0.09  \\
9:53:48.0 & 69:18:09 & 07.4 & 0.05  \\
9:53:49.1 & 69:17:55 & 09.0 & 0.06  \\
9:53:49.1 & 69:17:35 & 09.0 & 0.06  \\
\hline
\end{tabular}
\end{center}
\end{table}

The molecular complex is found to coincide with a maximum of H\,{\sc i} emission (see Fig.\,\ref{fig:1}). This suggests a recent formation  by condensation of the dense atomic gas in which it is embedded. It can therefore  be expected that neighbouring regions of similar column density harbour similar complexes. To test this scenario, a survey of CO has been undertaken around the position of the complex, and in the nearby H\,{\sc i} concentrations called ``Concentrations\,{\sc i} and {\sc ii}'' in \citet{1975ApJ...195...23G}. This survey of 125 pointings has been carried out with the IRAM 30\,m and the Kitt Peak telescopes (see Fig.\,\ref{fig:1}). None of the observations has given any detection suggesting that the molecular complex is unique in this region.  The Kitt Peak 115\,GHz observations smoothed to 5\,km\,s$^{-1}$ resolution have an rms in the range 6-10\,mK on a $T^*_{\rm R}$ scale and the 230\,GHz observations smoothed to 4\,km\,s$^{-1}$ have an rms in the range 8-14\,mK.  The results of the 30\,m observations are given in Table \ref{tab30m}. The corresponding 3\,$\sigma$ upper limits for the H$_2$ column density are close to $10^{20}$\,cm$^{-2}$ for a standard conversion factor as given in Sect.\,\ref{sec:mass}.

The  H\,{\sc i} concentrations surveyed are large and it appears unlikely that the  critical density can be reached at one location only if the physical conditions are similar throughout the  H\,{\sc i} concentrations. Instead, the uniqueness of the complex suggests the presence of a local peculiar phenomenon responsible for creating the physical conditions required to its formation.

\subsection{Situation of the complex in the large scale dynamics}\label{sub:dyn}

\begin{figure}
\centering
\resizebox{\hsize}{!}{\rotatebox{0}{\includegraphics{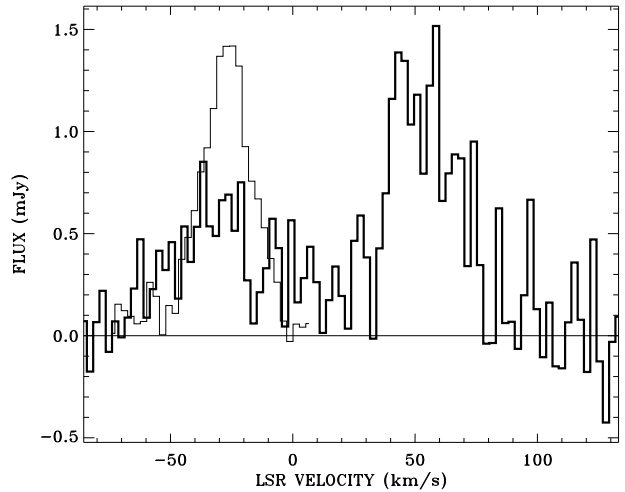}}}
\caption{H\,{\sc i} spectrum (thick line) at the position of the complex from the data of \cite{1996AJ....111..735A} spatially smoothed (beamsize: 18$''$). The CO(1--0) spectrum of the data inside the 4\,$\sigma$ level of the integrated map (Fig.\,\ref{fig:maps}) is superimposed (thin line). To match the H\,{\sc i} flux range the CO flux was divided by 700.}
\label{fig:HIspec}       
\end{figure}

\begin{figure*}
\centering
\resizebox{\hsize}{!}{\rotatebox{0}{\includegraphics{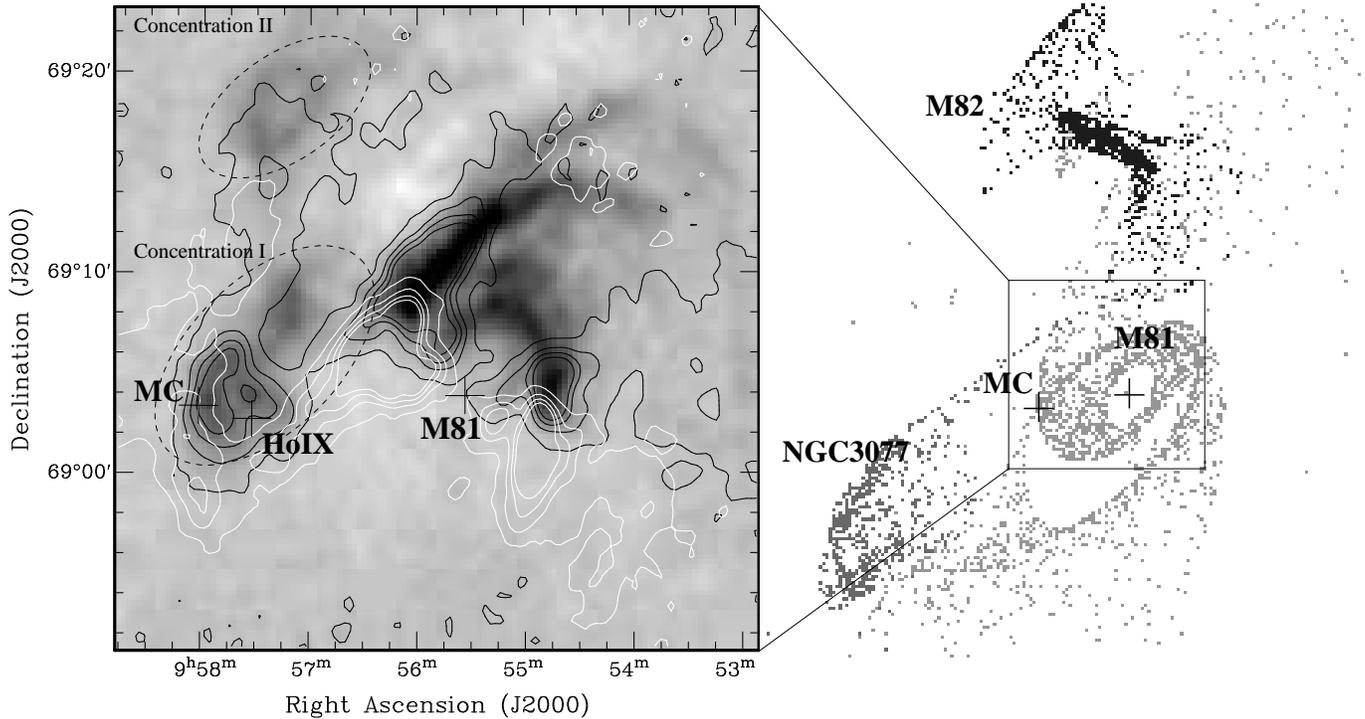}}}
\caption{Left:  H\,{\sc i} emission integrated over $[-132,239]$ km\,s$^{-1}$ from \cite{1994Natur.372..530Y} overlaid with contours of the same emission integrated over $[-80,0]$ km\,s$^{-1}$ (white) and $[0,80]$ km\,s$^{-1}$ (black). The levels are respectively $(0.15, 0.55, 0.95, 1.35)$ Jy/beam\,km\,s$^{-1}$ and $(0.1, 0.7, 1.3, 1.9)$ Jy/beam\,km\,s$^{-1}$. The crosses show the positions of the molecular complex, Holmberg\,IX, and the center of M\,81. Right: simulations of the interaction of M\,81  with M\,82 and NGC 3077 from \citet{1993egte.conf..253Y}.}
\label{fig:m81dyn}       
\end{figure*}

Understanding the situation of the complex with respect to the dynamics of the M\,81 group of interacting galaxies can help understanding its origin and properties. Is the complex at the interface of two  H\,{\sc i} structures? Where is it located exactly: is it intergalactic as was proposed in previous papers, is it in the disk of M\,81, is it in Holmberg\,IX? These are open questions that we address in this subsection.

As already noted in \citetalias{1992A&A...262L...5B}, H\,{\sc i} observations  show that the complex could be located at the boundary of two  H\,{\sc i} entities of different velocities. Indeed the H\,{\sc i} spectrum at the position of the complex (Fig.\,\ref{fig:HIspec}) consists of two distinct velocity components centered at $V_{\rm LSR}\sim -35$\,km\,s$^{-1}$ and  $V_{\rm LSR}\sim 50$\,km\,s$^{-1}$, and from their spatial distribution (Fig.\,\ref{fig:m81dyn}) each one appears to be prominent on one side of the complex.
 The first component is more prominent to the East and the second component, corresponding to the Concentration\,{\sc i}, is more prominent to the West. Are these two components cospatial (i.e. in physical contact or very close to each orther) or is it just a projection effect? Where are they located with respect to the disk of M\,81?

 In order to answer to this question we investigated the N-body simulations of the M\,81 group of galaxies from \citet{1993egte.conf..253Y}. Though these simulations are not supposed to reproduce exactly the morphology and the velocities at small scale, they can be used as a guide when trying to deproject the observations. The simulations show two components overlapping at the location of the complex as well as an overdensity at the edge of the disk of M\,81 similar to Concentration\,{\sc i} (Fig.\,\ref{fig:m81dyn}).  Rotating the cube of particles we find that it cannot be excluded that the two  H\,{\sc i} components are cospatial.  Indeed, some particles with the two velocities are found to be cospatial in the simulated 3-D cube at the edge of the perturbed disk of M\,81 and at the beginning of the tidal bridge toward M\,82. This is consistent with the H\,{\sc i} observations of \citet{1994Natur.372..530Y} which suggested that Concentration\,{\sc i} is inside the disk of M\,81. It would have resulted from the collision with M\,82 about $2\times 10^8$ years earlier.

What does it imply for the complex? The velocity of the complex is consistent with the first  H\,{\sc i} component  ($ -35$\,km\,s$^{-1}$). It could therefore be cospatial to this component which we just showed could itself be cospatial to Concentration\,{\sc i} (the second  H\,{\sc i} component) and be inside the extreme outer disk of M\,81. Thus, despite the velocity difference, it is not excluded that the molecular complex is inside or very close to Concentration\,{\sc i}  and in the disk of M\,81.

Holmberg\,IX, having a velocity of $V_{\rm LSR}\sim 125$\,km\,s$^{-1}$, may neither be related to the complex (and therefore the first  H\,{\sc i} component), nor to Concentration\,{\sc i} (and therefore the second  H\,{\sc i} component). It is moreover spatially offset by  $\sim$$2\ffam 8$ ($\sim$2.9\,kpc) from both objects.

The complex is at $13\ffam 3$ from the nucleus of M\,81. Therefore, if it is inside the disk of M\,81, it must be at a galactocentric radius of 24.2\,kpc assuming a distance of 3.6\,Mpc, a disk inclination of 59\,$\deg$ and a position angle of $149\deg$ for M\,81.  This would  be the most massive molecular association found in the extreme outer disk of a galaxy but not the most distant one: less massive molecular clouds are found up to 28\,kpc in our own galaxy \citep{1994ApJ...422...92D}. 
We note that for extreme outer regions of perturbed disks the boundary between 'galactic' and 'intergalactic' cannot be clearly defined. Also, not having a clear understanding of the dynamics in this region, a distinction between galactic and intergalactic is difficult to make.


\subsection{The nearby supershell and X-ray source}\label{sec:shell}

\subsubsection{General properties}

First identified by \citet{1988ApJ...325..544F}, X-9 is an ultraluminous X-ray source (ULX) situated $40''$ N-W of the complex. At the distance of M\,81 its X-ray luminosity is $10^{40}$\,erg\,s$^{-1}$, i.e. about half that of the nucleus of M\,81. \citet{1995ApJ...446L..75M} identified the  extended emission already observed by  \citet{1993A&A...273L..15H} to be a bright H$\alpha$ shell, ``MH9/10''\footnote{MH9/10 was initially thought to consist of two distinct H\,{\sc ii} regions: MH9 and MH10 in \citet{Miller94}, hence the name.}, of 250\,pc by 475\,pc in size (at the distance of M\,81) coincident with X-9. The large dimension of the shell -- comparable to  the size of supershells rather than supernova remnants -- and the high X-ray luminosity of X-9 make the nature of these two associated objects puzzling. A possibility would be that both objects are in the Galaxy and that X-9 is a runaway accreting white dwarf. But X-ray spectral and variability analysis argue against it and rather suggest an extragalactic accreting Kerr black hole binary \citep{2001ApJ...556...47L,2002MNRAS.332..764W}. According to the velocity of the H$\alpha$ emission ($V_{\rm LSR}\sim$56\,km\,s$^{-1}$) the supershell could actually be located inside Concentration\,{\sc i}. The dynamical timescale of the shell $\sim 1.5\times v_{{\rm s}2}^{-1} \times 10^{6}$\,yr (where $v_{{\rm s}2}$ is the shock velocity in units of $10^2$\,km\,s$^{-1}$)   is consistent with this picture: it is younger than concentration\,{\sc i} resulting from the pass by of M\,82  $\sim 10^{8}$\,yr ago (see previous section).

If the radiation of the compact object is unbeamed, X-9 should be an intermediate mass black hole (IMBH) of $\sim$$100\, {\rm M}_{\odot}$;  otherwise it could be a stellar-mass object like a microquasar. The H$\alpha$ supershell could fit to either scenario: it could have formed during a hypernova explosion at the collapse of the IMBH or being powered by an intense outflow from a microquasar. Few other ULX associated with nebulae have been observed and for all of them it remains difficult to choose between these scenarios \citep{2003RMxAC..15..197P,2003MNRAS.342..709R}. The roundish shape of MH9/10 seems to argue against the outflow hypothesis but 20~cm radio continuum observations of  \citet{1986ApJ...310..621B} suggest non-thermal emission. Also some blue emission at the center of the supershell described as a 'blue knot' is coincident with  X-9 \citep{1995ApJ...446L..75M} and could well be the signature of an outflow rather than that of an accreting disk. Recent XMM-Newton observations lend support to the IMBH hypothesis  \citep{2004ApJ...607..931M}.

\subsubsection{Description of the new optical observations}

The new H$\alpha$ observations (Fig.\,\ref{fig:Ha}) show for the first time that  the emission at the S-E of the supershell, known as MH11, is actually connected to the supershell and is as large as the supershell itself. This emission appears as an extension of the supershell in the direction of the complex and  covers about half of the distance to it i.e. $\sim$20'' ($\sim$350\,pc). It appears to be composed of two parallel ridges elongated in the direction of the molecular complex with some lower level emission in between. 

The B-band image  (Fig.\,\ref{fig:B}) shows that the 'blue knot' (B1) inside the supershell mentioned by \citet{1995ApJ...446L..75M} is actually of elongated shape oriented in the E-W direction. Most remarkably its Western end coincides with the Chandra position of X-9, i.e. $\alpha_{J2000}=09^h 57^m 53\ffs 3$, $\delta_{J2000}= 69^{\circ}03'46\ffas 4$ \citep[from Chandra archival data presented in][]{2002AAS...200.0810E}. The Chandra absolute position accuracy is better than $1''$ and we can conservatively estimate that our astrometric error does not exceed 2$''$. Then the uncertainty is less than $3''$ which corresponds to the size of the cross on Figs.\,5--7.  On its Eastern side the elongated emission  seems to end in front of a small structure (B2) coinciding with the inner boundary of the H$\alpha$ supershell. 
Another small blue structure (B3) can be seen on the other side of the supershell boundary (on the external side) but slightly more to the South i.e. at the beginning of the H$\alpha$ extension.

All these features can also be seen at a lower level in the R-band frame (Fig.\,\ref{fig:R}). Some granularity can be seen in the  R-band emission extending from X-9.

\subsubsection{Possible interpretations}\label{sec:interp}

The H$\alpha$ morphology suggests that the extension could be a large outflow from the supershell. The supershell would have poured out its inner high pressure medium into a medium with lower density, according to the well known champagne effect. A similar blowout  with a radial velocity of 90\,km\,s$^{-1}$ and from a superbubble of 60\,pc in diameter was observed in the H\,{\sc ii} complex N44 of the LMC by \citet{1996ApJ...464..829M}.  We lack the velocity information to check any gradient along the H$\alpha$  extension.  From the geometry we can however compare the energetics of such a large outflow with that of the supershell. If the supershell is assumed to be a sphere of radius $R_{\rm s}$ with kinetic energy $E_{\rm s}$  and the outflow is a cylinder of same radius, of length $L_{\rm o} (\sin{i})^{-1}$ ($i$ being the angle of the outflow with the line of sight) and of kinetic energy $E_{\rm o}$, then the ratio of the energies is given by:
\begin{equation}\label{eq:energies}
\frac{E_{\rm o}}{E_{\rm s}}=\frac{3}{4}\frac{L_{\rm o}}{R_{\rm s} \sin{i}}\alpha\beta^2,
\end{equation} 
where $\alpha$ is the ratio of the average initial density in the region of the outflow to the preshock density of the supershell, and $\beta$ is the ratio of the average velocity in the outflow to the velocity of the shock of the supershell. We have $R_{\rm s}$$\sim$$150$\,pc, $L_{\rm o}$$\sim$$150$\,pc. The orientation of the outflow is not known but for $i$$>$$15\deg$, $\sin{i}$$>$$0.25$, then $E_{\rm o}/E_{\rm s}<7\times\alpha\beta^2$. If the outflow is due to the champagne effect the initial density must be lower in the region of the outflow, i.e. $\alpha<1$, and the gas is accelerated further in the blowout but averaged over the whole outflow the velocity is expected to be comparable to the shock velocity of the shell i.e.  $\beta\gtrsim 1$. A variation by a factor 10 in density is possible, i.e. $\alpha=0.1$ is possible, and therefore $E_{\rm o}\lesssim E_{\rm s}$ cannot be excluded. We conclude that, considering its extension, the energetics of such a blowout is not inconsistent with that of the supershell.

The blue structure B1 (the 'blue knot') could be a jet powered by X-9. On the plane of the sky it appears oriented in a direction $45\deg$ away from that of the ouflow but this might be a projection effect and both might be better aligned. Then this hypothetical jet could inject kinetic energy  into the outflow. The orientation could also have changed if the jet is precessing like in SS\,433 \citep[e.g.][]{1982Sci...215..247M}. Even if misaligned with the extension the jet could  be responsible for the formation of the supershell and for maintaining the high inner pressure feeding the large outflow. B2 (Fig.\,\ref{fig:B}) could be the impact of the jet onto the shell and the granularity of the red emission could correspond to knots in the jet. 

Other interpretations of the H$\alpha$ extension and the blue features are however possible. The H$\alpha$ extension could result from X-ray photoionisation rather than a blowout. Such a process was proposed for the  H$\alpha$ emission of 800\,pc in extent near the ULX NGC\,1313 X-1 \citep{pakull2002}. Also, B1, B2 and B3 could be associations of young stars similar to those seen around the shell. Thus, B1 could well be a central cluster and B2 and B3 could result from star formation inside the wall of the supershell.  For the hypothetical jet from the X-ray source high resolution  radio continuum observations would be useful. Deeper  H\,{\sc i} interferometric observations might also help to check the dynamics of the gas at the location of the supershell.

\section{Discussion}\label{sec:disc}

Could the complex have formed at the interface of the two  H\,{\sc i}  components discussed in Sect.\,\ref{sub:dyn}? Such a scenario was proposed for the cometary cloud G110-13 by \citet{1992ApJ...397..174O}. This hypothesis leads, however, to a cloud that is elongated along the interface, i.e. in the N-S direction (Fig.\,\ref{fig:m81dyn}). Instead the complex is elongated toward the S-E, $>$45\,$\deg$ away from the interface orientation.

Is the molecular complex shaped by an outflow from MH9/10?
It was noted in Sect.\ref{sec:morph} that, according to the CO emission, the molecular complex had a  filamentary/cometary morphology with the head oriented toward the N-W. Then, the optical observations of the supershell situated on the N-W of the complex revealed an H$\alpha$  extension toward the complex that could be interpreted as an outflow (Sect.\,\ref{sec:shell}). The apparent coincidence of these two observations lead us to consider the possibility that the complex is shaped by the hypothetical outflow from MH9/10.

 Such filamentary/cometary  shapes have been observed in molecular clouds for years, e.g. in the Orion cloud \citep{1987ApJ...312L..45B} with a mass of $\sim 5\times 10^4~{\rm M}_{\odot}$ and a size of 13~pc, the Draco cloud \citep{1987ApJ...318..702O} with a mass of $\sim 150~{\rm M}_{\odot}$ and a size of $\sim 10$~pc and at galactic scales  in the Magellanic Clouds \citep[e.g.][]{1977ApJ...217L...5M}. A sample of 15 comet-like clouds selected in the IRAS survey was studied by \citet{1988ApJ...325..320O}.
One of the scenarios invoked  to account for these morphologies involves interaction of the clouds with the surrounding interstellar medium.

In the present case, if the outflow is interacting with the complex it has to extend over $\sim$$50''$, i.e. it must be 2.5 times larger than the H$\alpha$ extension. Its energy can be compared to that of the supershell in the same way as done in the previous section (Eq.\,\ref{eq:energies}). Assuming the cometary shape is oriented like the outflow, the distribution of CO in the data cube can help to estimate the $\sin i$ parameter, i.e. the angle between the outflow and the line of sight. To that purpose we fit the CO(1--0) data cube with a simple cometary model consisting of a cone in which the density decreases exponentially with the distance from the head and the modulus of the velocity increases linearly. Such a conical shape is suggested by the way the  clumps  and the filaments discussed in Sect.\,\ref{sec:morph} are distributed (Fig.\,\ref{fig:chanmaps}). The distribution of the emission in the channel maps  changes almost monotonically from compact at $-$10~km\,s$^{-1}$, to extended or spread mostly on the S-E side of the compact emission at $-$35~km\,s$^{-1}$.  This can be interpreted as a cometary shape with the head pointing to the N-W and away from the observer and a tail expanding toward the S-E and the observer. The best fit of the cone model gives  $i$$\sim$$30 \deg$. The channel maps of the model are overlaid in contours to the CO(1--0) in Fig.\,\ref{fig:chanmaps} and its integrated intensity is shown on the right-hand side of Fig.\,\ref{fig:maps}. While it does not reproduce correctly the details of the individual CO(1--0) channel maps, it reproduces the general trend noted above. It is interesting to note here that this interpretation of the morphology and dynamics of the CO emission would provide an explanation to the discrepancy by a factor of 10 between the virial mass and the molecular mass determined from the CO line intensity noted in \citetalias{1992A&A...262L...5B}, as the complex would then be gravitationally unbound.
 Then, according to the calculations of the previous section the kinetic energy of this outflow relates to that of the supershell as: $E_{\rm o} \sim 10 \times \alpha\beta^2 E_{\rm s}$. Hence for this outflow to be realistic we must have $\alpha\beta^2<0.1$. As discussed in Sect.\,\ref{sec:interp}, this is  not impossible,  e.g. $\alpha=1/50$ and $\beta=2$. The size of the  outflow would then be $\sim$1.8\,kpc which is larger than any shell blowout observed so far.

In this hypothesis the complex could result from the compression of a higher density cloud embedded in the low density region of the outflow, with an initial line of sight LSR velocity $\gtrsim -10$\,km\,s$^{-1}$ (velocity of the head of the cometary shape).  This scenario would then provide an explanation to the isolation of the complex (Sect.\,\ref{sec:env}). For molecular gas to be seen at such a large galactocentric radius, peculiar physical conditions might be required which this unusual supershell outflow would provide. Conversely it was already noted in \citet{1995ApJ...446L..75M} and \citet{2002MNRAS.332..764W} that X-9 and the nearby stars are young objects which formed out of molecular gas and therefore one expects to find molecular gas around. The complex may not result only from the condensation of the compressed H\,{\sc i} but also from the concentration of pre-existing molecular gas under the action of the outflow.

Following this scenario, star formation might occur inside the head of the cometary complex, that is however not seen. Spectral observations of the H$\alpha$ extension would be useful, because these provide information on the kinematics of the ionized gas as it points toward the molecular complex. Deeper H$\alpha$ imaging is also needed to check how close the S-E extension of the ionized gas gets to the molecular complex.

\section{Conclusions}
The study of the properties and the environment of a molecular complex of mass 2--6 $\times$ 10$^6$\,M$_{\odot}$, $\sim$13$'$ east of M\,81, leads to the following main results:'

\begin{enumerate}
\item We confirm there is no unobscured massive star formation inside the complex (SFR$<$$10^{-6}\,{\rm M}_{\odot}$\,yr$^{-1}$ for a Salpeter IMF). Furthermore, the upper limit on the extinction ($A_{\rm V}$$<$$5.5$) imposed by the millimeter continuum observations implies that the SFR, even obscured, must be lower than expected for such a mass of molecular gas  (SFR$<$$10^{-4}\,{\rm M}_{\odot}$\,yr$^{-1}$ for a Salpeter IMF). 

\item The CO line ratios admit a solution corresponding to abundances and physical conditions similar to those found in the disks of spiral galaxies.

\item We find from its dynamics (no rotation) and its mass (2--6\,$\times 10^6 \,{\rm M}_{\odot}$) that the complex resembles a massive GMC (or a GMA) rather than a dwarf galaxy. 

\item From the inspection of N-body simulations of the M\,81 group and the H\,{\sc i} data we find that it might be located inside the extreme outer disk of M\,81 and be cospatial with the H\,{\sc i} feature known as Concentration\,{\sc i}. Thus, instead of being inside a well detached tidal arm it could be located  in a region of the outer disk of M\,81 which was strongly perturbed by the interaction with M\,82. 
\item The negative results of the CO survey in the nearby H\,{\sc i} condensations suggests that the complex is unique in this region. This calls for a peculiar local formation process.
\item More tentatively we note that the distribution of the CO emission in the data cube is asymmetric in a way consistent with a cometary object pointing away from the observer and toward the N-W. We propose that the complex is shaped by interaction with a large outflow from the nearby supershell MH9/10. It could also have formed by compression of the ISM under the action of the outflow.  New observations are required to test this hypothesis.
\end{enumerate}

Points 4 and 5 imply that  this complex could be different in nature and origin from the one found by \citet{1999ApJ...519L..69W} near NGC\,3077. The latter  coincides with an H\,{\sc i} peak inside a tidal arm, it is clearly intergalactic and has most probably formed by condensation of the highest density H\,{\sc i} concentration of the arm. These differences could provide an explanation for the different star forming properties of the two complexes: \citet{2003IAUS..217E.158W} reported an SFR of $2.6\times 10^{-3}\,{\rm M}_{\odot}$\,yr$^{-1}$ in the complex near NGC\,3077.

\acknowledgement
We are greatful to our referee, Ute Lisenfeld, for comments  that helped to significantly improve the quality of the paper. We acknoledge M.\,Yun and D.S.\,Adler for having kindly provided their H\,{\sc i} data several years ago. We also thank F.\,Gueth for providing the SHORT\_SPACE task before it was included in the GILDAS software. We used the Karma software developed by \citet{1995adass...4..144G}. We also thank the IRAM staff of Grenoble and Granada and the NRAO staff of Tucson/Kitt Peak. Research by F.\,B. was supported by a postdoctoral fellowship from the Universities of Bochum and Bonn in the frame of the Graduiertenkolleg 787.

\bibliographystyle{aa}
\bibliography{imc}

\end{document}